# VAT Compliance Incentives[*]

Maria-Augusta Miceli [†]

*Dept. of Economics and Law*
*Sapienza University of Rome*

February 25, 2021

**Abstract**

VAT evasion is a collusive behavior between the two counterparts in a transaction, jointly interacting with the Tax Authority. We first check the tax parameters for Buyer and Seller to agree on complying or evading, under different tax policy scenarios: (i) normal taxes, (ii) tax allowances. Then we check if, how and for which fiscal parameter thresholds, compliance resists, in a Bayesian game, against a Tax-Authority auditing with a random probability.

Fiscal parameter thresholds are designed to make compliance a dominant strategy even under No-Audit. In particular thresholds are not only functions of audit probability or tax allowances, but also of the cost structure of the firm.

---

[*]I do thank Silvia Fedeli for early discussions and the many pratctioneers I met along the way.

[†]Address: Dept. of Economics and Law, Sapienza University of Rome, via del Castro Laurenziano 9, 00161 Rome, Italy. Email: augusta.miceli@uniroma1.it.

# 1 Introduction

VAT evasion is a strategic collusive behavior between the two counterparts in a transaction, the Buyer and the Seller, jointly interacting with the Tax Authority. We first check the tax parameters allowing Buyer and Seller to agree on the same strategy, complying or evading, under different tax policy scenarios: (i) normal taxes, (ii) tax allowances on specific VAT expenses. By agreeing, they form a coalition. Then we check whether, how and for which fiscal parameter thresholds, agents' coalition compliance resists, in a Bayesian game, against a Tax Authority auditing with a random probability.

Fiscal parameter thresholds are designed to make compliance a dominant strategy even under No-Audit. In particular, thresholds are not only functions of the existence of audit probability or tax allowances, but also of the cost structure of the firm.

Traditionally models of tax compliance deal with the strategic interactions between the tax payer and the tax authority about income tax. The peculiar eInitial historical results were linking evasion to the tax payer degree of risk aversion with respect to audit and sanctions (Allingham and Sandmo (1972), Sandmo, A. (1974), Cowell (1985), (1990))). Some models justify compliance by including tax morale issues (Erard, B., & Feinstein, J. S. (1994), Bénabou, R., & Tirole, J. (2006), Torgler (2003), (2007), Luttmer and Singhal (2014), Lisi (2015)) or general equilibrium ones (pay all to pay less).

Even under defined game theory approaches (Reinganum and Wilde (1985), Graetz, Reinganum and Wilde (1986), Cronshaw and Alm (1995), Carfì, D., & Musolino, F. (2015)) the peculiar relation from indirect to income taxes is neglected. Others use experiments to learn probabilities from sequential behaviors (for example, Graetz, Reinganum and Wilde (1986), Alm et al. (1992), Alm and McKee (2004) and Bloomquist (2011)).

Here we use the specificity of the cross evasion from VAT to the involved income taxes to get the mentioned thresholds on tax parameters. We explicitly avoid modelling evasion as linked to the variable listed above: the degree of risk aversion, the general equilibrium expectations, tax morale utilities and, of course, the absence of audit. We instead assume agents to be selfish and risk neutral. The probability of audit is modelled as an endogenous variable, chosen by the Tax Authority to obtain compliance as Bayesian game dominant strategy, given the tax rate parameters.

To simplify the analysis, the attention is focused on the final transaction of a production chain between the final registered Seller and the Buyer, where the value of the transaction and its VAT are parametrically defined. While under no audit, Buyer and Seller generally agree on complying although with different degrees of preference, as audit probability increases, both switch to evasion, but at different values of that probability, depending on the importance of the cost structure and fiscal tax rates. This is the issue explored.

That tax allowances help compliance is a known procedure and at work in many countries. This model tries to fix a numerical methodology to empirically compute the numerical thresholds for the tax parameters, able to lead to the results.

We start with defining the notation, the strategies for each player, the fiscal scenarios and the procedure. Then we set the game between the two private counterparts, we show the payoffs and highlight the parameter boundaries for compliance to be the dominant strategy. Once learned the conditions for the private coalition to be formed, we build the Bayesian game between the private coalition and the Tax Authority, to study how allowance policy changes the audit frequency needed for compliance, as the cost structure of the firm and the fiscal rates change.

# 2 Model Settings

The analysis is at micro level and considers three agents: two private ones, the registered trader or "seller" (S) of the final good and the consumer or final "buyer" (B) on which the whole VAT accrues, and the government tax authority (G) that collects income tax and VAT revenue.

We focus on evasion happening on the only final transaction of a final good or service between the two private counterparts, the final seller and the final buyer. Each of the two agents has a fully declared income, prior to the transaction, respectively $y_S$ and $y_B$, that determines their initial marginal tax revenue rate.

## 2.1 Definitions and Notations

- Players:
  - The registered trader or "seller" (S)
  - The final "buyer" (B)
  - The government tax authority (G)

- $y_S, y_B$ = private agent incomes.

- $(x_O - x_I)$ = the last transaction between the final registered seller and the buyer, subject to VAT, where $x_O$ = output value, $x_I$ = input value.

- $Y_B, Y_S, Y_G$ = net income after the transaction, function of joint agent strategies.

- $t_S, t_B$ = marginal income tax rates, that would not change because of the transaction.

- $v = VAT$ rate.

- $s_y, s_v$ = sanction rates on, respectively, income and VAT tax.

- $\theta$ = deduction rate, i.e. percentage of selected invoiced expenses that can be deducted from buyer income tax .

- $\gamma$ = audit probability.

- Parameters for the plots: $x_0 = 1$, $t_S = 0.24$, $v = 0.22$, $\delta = 1$; while $x_I \in [0,1], \theta \in [0,1], \gamma \in [0,1]$, unless fixed to some values to simplify graphs.

## 2.2 Simplifying assumptions

1. Focus on agents tax compliance on the last transaction of the production chain $(x_O - x_I)$, representing the surplus to redistribute.

2. Since we only want to deal with the opportunity to evade income and VAT tax on a single transaction, we assume that each agent is compliant over the rest of his income, i.e. the income before the transaction has been regularly declared by each agent: $y_S^d = y_S$ and $y_B^d = y_B$

3. Marginal income tax rates for the seller $S$ and for the buyer $B$ do not change because of the transaction.

4. To be realistic, although evaded VAT should stand on buyer responsibility, it is almost impossible to catch him, so we assume that omitted VAT invoice is sanctioned on the final seller.

## 2.3 Strategies for each player

**Buyer (B)**

1. $\mathbf{C_B}$ = **Comply**
2. $\mathbf{E_B}$ = **Evade**

**Seller (S)**

1. **$C_B$ = Comply**

2. **$E\_LT1_S$ = Evade just the Last Transaction, but deducing income costs.**
   - Costs $x_I$ are deductible from imposable income.
   - VAT on inputs is paid by the final seller and cannot be translated.

   The idea is that the seller acts as final buyer on some costs, he is able to deduce costs, but unable to pass on VAT. Essentially he not only evades income tax on $x_O$ revenue, but he also reduces net income by deducting the costs that he imputes to previous revenues. This type of evasion is possible when the seller has many last transactions and can impute costs to another, declared one.

3. **$E\_LT2_S$ = Evade just the Last Transaction**
   - Costs $x_I$ are "not" deductible from imposable income.
   - VAT on inputs is paid by the final seller and cannot be translated.

4. **$E\_WT_S$ = Evade the Whole Transaction**:
   - Costs $x_I$ are "not" deductible from imposable income;
   - VAT has never been paid along the production chain.

**Tax Authority (G)**

1. Audit and Sanction (A) with probability $\gamma$,
2. No-Audit (NA) with probability $(1 - \gamma)$.

## 2.4 Tax Policy Scenarios

0. **No Taxes**, meaningless, but used as a benchmark. **(NT)**
1. **Taxes** with NO-deductions **(T)**
2. **Tax Deductions:** specific selected invoices can be deducted from the buyer income tax **(TD)**

# 3 Modelling Procedure

## 3.1 Game: Buyer vs Seller, for each Govt. Tax Policy

In the games with one buyer and one seller as counterparts of a unique transaction, the strategies should be agreed by both players. Therefore the off diagonal elements of the matrix do not exist. Moreover, the consumer, by definition participating only in the final transaction, can just evade that single transaction, while he is the counterpart of the many seller evasion strategies described above ($E\_LT1_S, E\_LT2_S, E\_WT_S$)

Let's consider the simple matrix form game (Game1), that will be set for each scenario.

| Buyer \ Seller | $C_S$ | $E\_LT1_S$ | $E\_LT2_S$ | $E\_WT_S$ |
|---|---|---|---|---|
| $C_B$ | (1) | $\nexists$ | $\nexists$ | $\nexists$ |
| $E_B$ | $\nexists$ | (2) | (3) | (4) |

((Game 1))

Each cell represents the event determined by the joint strategies due to each agent, the buyer, the seller and the Tax-Authority, showing the three payoffs. The third agent, the Tax Authority (G, form now on) will have the same strategy (Audit or NoAudit) on the entire matrix.

Since the counterparts of a same transaction must both comply or both evade, only four cases are meaningful. While Evading, though, the seller must choose whether evading just the last transaction and whether deducing costs (E_LT1) or not (E_LT2), or evading the total production chain (E_WT)

This game is studied under the above three "tax policy scenarios": No Taxes (NT), Taxes (T), Tax Deductions (TD).

Each game under the three scenarios is repeated again for each Tax Authorities strategy:

1. No Audit $(\gamma = 0)$;

2. Audit and Sanctions $(\gamma = 1)$;

3. Uncertain Audit and Sanctions (Bayesian Game BG) : $\gamma \in (0,1)$., which is simply the expected value of the two previous scenario using the probability $\gamma$.

### 3.2 Solution Methods

In order to find a solution we proceed in two stages.

#### 3.2.1 Stage 1

| Buyer \ Seller | $C_S$ | $E\_LT1_S$ | $E\_LT2_S$ | $E\_WT_S$ |
|---|---|---|---|---|
| $C_B$ | (1) | $\nexists$ | $\nexists$ | $\nexists$ |
| $E_B$ | $\nexists$ | (2) | (3) | (4) |

(1)

In each of these games, we look for strict dominant preferences towards compliance for both private agents, against the three possible types of evasion. When no strict dominance exist, compliance, is studied as a function of tax parameters and sanction rates. As obvious, agent behaviour changes in the different policy scenarios:

- while in absence or presence of controls (probability of audit) the buyer generally follows the seller preference,

- in presence of tax deductions (TD), the buyer may prefer to comply, even facing a seller wishing to evade. The degree of preference is always determined by the tax parameters.

#### 3.2.2 Stage 2

We consider the payoffs of the joint coalition "Buyer + Seller". The coalition plays against the Tax Authority, whose strategy is to Audit with probability $\gamma$.

This game is peculiar since the private coalition has the choice between four discrete strategies, while the Tax Authority should choose his best response to each strategy by choosing $\gamma$, which is a continuous choice.

| \ G | Tax-Policy-Scenario |
|---|---|
| (B+S) | $\gamma$ |
| $C_B + C_S$ | ... |
| $E_B + E\_LT1_S$ | ... |
| $E_B + E\_LT2_S$ | ... |
| $E_B + E\_WT_S$ | ... |

(2)

The above game is repeated for each of the two Tax Policy scenarios (Normal taxes (T) and Taxes with deductions (TD)).

The computation of the thresholds to achieve strategic dominance of compliance is the main result of this work.

Results have different nuances along the scenarios, so all the details are shown for the No-Audit case, while afterwards we will only underline the important differences.

## 3.3 Scenario 0. No-Taxes: $NT$

The No-Taxes scenario is written just to fix an initial grammar and establish the benchmark.

In absence of taxes, the existence of the Tax Authority is irrelevant and the strategies "Comply" (C) and "Evade" (E) are equivalent, therefore payoffs in cells (1), (2), (3) and (4) in Game 1 are all the same for any audit probability $\gamma \in [0,1]$.

The payoffs in each cell by each agent are as follows

$$\{C_B, C_S\} = \{E_B, E\_LT1_S\} = \{E_B, E\_LT2_S\} = \{E_B, E\_WT_S\}:$$
$$\begin{cases} Y_B = y_B - x_O \\ Y_S = y_S + x_O - x_I \\ Y_G = 0 \end{cases}$$

Moreover

**Definition 1** *In all events, the sum of the net incomes after the transaction will always be defined as the sum of agent incomes pre-transaction, minus the costs due to the previous segment of the production chain. The government has no previous income. This represent the economic surplus to be distributed*

$$Y_B + Y_S + Y_G = y_B + y_S - x_I \qquad (3)$$

Total net income as defined above in eq. (3) must hold in each game cell for all the games presented below.

# 4 Scenarios A: No-Audit($\gamma = 0$)

Let's start under the assumption that the tax authority doesn't audit $(\gamma = 0)$, associated with each of the tax policy scenario. The aim of this section is to determine the parameter conditions leading to Nash Equilibria of each game.

## 4.1 Scenario A1. Taxes with no deductions under No-Audit: $(\gamma = 0)\_\&\_T$

### 4.1.1 Payoffs $(\gamma = 0)\_\&\_T$

Let's consider the matrix form of Game 1 (1), we describe in Table 1 below, the payoffs of each significant case.

1. Event "Compliance" $\{C_B, C_S\}$

2. Event "Evasion just the Last Transaction, 1" $\{E_B, E\_LT1_S\}$.

    The seller avoids income tax on the additional $x_O$, he pays all costs, but is unable to pass on VAT.

3. Event "Evasion just in the Last Transaction, 2" $\{E_B, E\_LT2_S\}$

    The seller avoids income tax on the additional $x_O$, he pays all costs, but is unable to reduce his net income nor pass on VAT.

4. Event "Evasion on Whole Transaction" $\{E_B, E\_WT_S\}$

    The seller avoids income tax on the additional $x_O$, but he cannot reduce his net income by deducting costs. VAT is totally evaded by both agents.

|  | Scenario A: NO-AUDIT ($\gamma = 0$) |
|---|---|
|  | Scenario A1: Taxes with No Deductions ($T$) |
| $\{C_B, C_S\}$ : | $\begin{cases} Y_B(C_B, C_S) = (1 - t_B) y_B - x_O(1 + v) \\ Y_S(C_B, C_S) = (1 - t_S)(y_S + x_O - x_I) \\ Y_G(C_B, C_S) = t_B y_B + t_S(y_S + x_O - x_I) + v x_O \end{cases}$ |
| $\{E_B, E\_LT1_S\}$ : | $\begin{cases} Y_B(E_B, E\_LT1_S) = (1 - t_B) y_B - x_O \\ Y_S(E_B, E\_LT1_S) = (1 - t_S)(y_S - x_I) + x_O - v x_I \\ Y_G(E_B, E\_LT1_S) = t_S(y_S - x_I) + t_B y_B + v x_I \end{cases}$ |
| $\{E_B, E\_LT2_S\}$ : | $\begin{cases} Y_B(E_B, E\_LT2_S) = (1 - t_B) y_B - x_O \\ Y_S(E_B, E\_LT2_S) = (1 - t_S) y_S + x_O - x_I(1 + v) \\ Y_G(E_B, E\_LT2_S) = t_B y_B + t_S y_S + v x_I \end{cases}$ |
| $\{E_B, E\_WT_S\}$ : | $\begin{cases} Y_B(C_B, C_S) = (1 - t_B) y_B - x_O \\ Y_B(C_B, C_S) = (1 - t_S) y_S + x_O - x_I \\ Y_G(C_B, C_S) = t_S y_S + t_B y_B \end{cases}$ |

To be able to plot the strategies in a quite general setting, we standardize the output value to $x_0 = 1$, and we consider $x_I$ as a percentage of the output value. to understand how the relative magnitude of the costs affects the evasion decisions.

Plot parameters: $y_S = 0$, $t_S = 0.24$, $v = 0.22$.

We set $y_S = 0$ on purpose, to notice for which strategy and which cost percentage the net income can go negative.

$$\begin{cases} Y_S(C_B, C_S) & = (1 - t_S)(y_S + x_O - x_I) = (1 - 0.24)(1 + 1 - x_I) \\ Y_S(E_B, E\_LT1_S) & = (1 - t_S)(y_S - x_I) + x_O - v x_I = (1 - 0.24)(1 - x_I) + 1 - .22 \cdot x_I \\ Y_S(E_B, E\_LT2_S) & = (1 - t_S) y_S + x_O - x_I(1 + v) = (1 - 0.24) 1 + 1 - x_I \cdot 1.22 \\ Y_S(E_B, E\_WT_S) & = (1 - t_S) y_S + x_O - x_I = (1 - 0.24) 1 + 1 - x_I \end{cases}$$

**Fig. 1.**

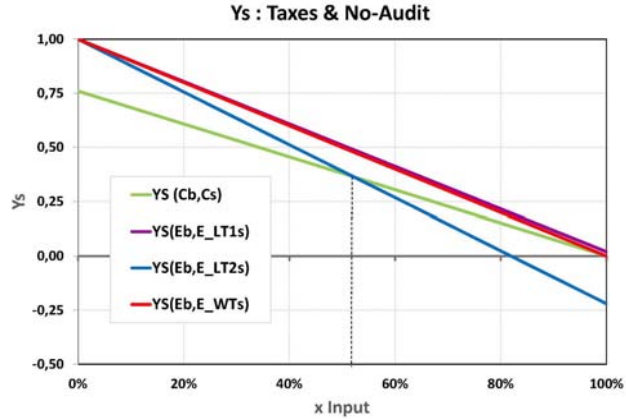

In fig.1. we plot seller net incomes $Y_S$ deriving from the four strategies $C_S$ (green line), $E\_LT1_S$ (burgundy), $E\_LT2_S$ (blue line) and $E\_WT_S$ (red line). The four strategies lead to the same net income $Y_S$ for costs $x_I = 0$. For any $x_I \in [0, 1]$, $E\_LT1_S$ and $E\_WT_S$ have similar magnitudes, because in $E\_LT1_S$ the seller reduces net imposable income with costs $x_I$, but he cannot pass on the VAT, while in $E\_WT_S$ the seller doesn't save on imposable income, but avoids VAT. Since in the Italy numerical example, these two rates are very close, the two strategies lead to similar outcomes. Essentially in $E\_LT1_S$ the seller compensate VAT compliance on costs by evading income tax. The evasion strategy $E\_LT2_S$ is the "most compliant" among the three evasion types, because the seller avoids income tax on $x_O$, but he is penalized by not deducting costs and being sunk on VAT.

**Remark 1** *Under No-Audit, we would expect that "evasion" would be in general more rewarding than "compliance". Instead, interestingly, over a certain cost threshold (that will be explained below in Corollary 1., compliance,$C_S$, is more profitable than the $E\_LT2_S$ strategy, because evading just $x_O$ impedes to reduce net income with costs.*

### 4.1.2 Strategic Dominance $(\gamma = 0)\_\&\_T$

Now we look for dominating strategies by comparing compliance with each of the evasion strategies. In this first case we show all the steps, while in the next sections, we will show just the results.

**Buyer** $(\gamma = 0)\,\&\,T$

$$\begin{aligned} Y_B(C_B) &\geq Y_B(E_B) \\ (1-t_B)y_B - x_O(1+v) &\geq (1-t_B)y_B - x_O \\ 0 &< vx_0 \\ &\Longrightarrow C_B \preceq E_B \end{aligned}$$

Under no audit, obviously evading means a gain of the VAT tax $(vx_O)$ for the Buyer.

**Seller** $(\gamma = 0)\,\&\,T$   1. $C_S$ vs $E\_LT1_S$

$$\begin{aligned} Y_S(C_S) &\geq Y_S(E\_LT1_S) \\ (1-t_S)(y_S + x_O - x_I) &\geq (1-t_S)(y_S - x_I) + x_O - vx_I \\ \underbrace{vx_I}_{\text{cost from evasion}} &\geq \underbrace{t_S x_O}_{\text{gain from evasion}} \\ &\Longrightarrow C_S \preceq E\_LT1_S \\ \text{for } t_S \leq \frac{vx_I}{x_O} \quad &\text{or} \quad v \geq \frac{t_S x_O}{x_I} \end{aligned} \tag{4}$$

**Lemma 1** $C_S \preceq E\_LT1_S$

**Proof.** Even if $x_I/x_O = 1$, $t_S > v$, compliance is never preferred to $E\_LT1_S$. Also, by construction, since $x_I \leq x_O$, then $vx_I \leq t_S x_O$ therefore the buyer will always evade. ∎

2. $C_S$ vs $E\_LT2_S$

$$\begin{aligned} Y_S(C_S) &\geq Y_S(E\_LT2_S) \\ (1-t_S)(y_S + x_O - x_I) &\geq (1-t_S)y_S + x_O - x_I(1+v) \\ \underbrace{vx_I}_{\text{cost from evasion}} &\geq \underbrace{t_S(x_O - x_I)}_{\text{gain from evasion}} \\ &\Longrightarrow C_S \succeq E\_LT2_S, \\ \text{for } t_S \leq \frac{vx_I}{(x_O - x_I)} \quad &\text{or} \quad v \geq \frac{t_S(x_O - x_I)}{x_I} \end{aligned} \tag{5}$$

**Lemma 2** $C_S \succeq E\_LT2$

**Proof.** Compliance is convenient if the "VAT reimbursement loss" is greater than "income tax saving". ∎

3. $C_{S\_}$ vs $E\_WT_S$

$$\begin{aligned}
Y_S(C_S) &\geq Y_S(E\_WT_S) \\
(1-t_S)(y_S + x_O - x_I) &\geq (1-t_S)y_S + x_O - x_I \\
\underbrace{0}_{\text{gain from compliance}} &\geq \underbrace{t_S(x_O - x_I)}_{\text{gain from evasion}} \\
&\implies C_S \precsim E\_LW_S, \qquad \forall t_S, v
\end{aligned} \qquad (6)$$

We can now summarize the thresholds under which the strategy "compliance" ($C_S$) would eventually be preferred versus the different ways to evade.

**Proposition 1** *Under the scenario Taxes & No-Audit* $(\gamma = 0)$, *the Seller would prefer to comply if*

$$t_S < \begin{cases} \frac{vx_I}{x_O} & \implies C_S \succeq E\_LT1_S \\ \frac{vx_I}{(x_O - x_I)} & \implies C_S \succeq E\_LT2_S \\ 0 & \implies C_S \preceq E\_WT_S \end{cases}$$

**Proof.** By $(4), (5)$ and $(6)$. ∎

**Fig. 2.**

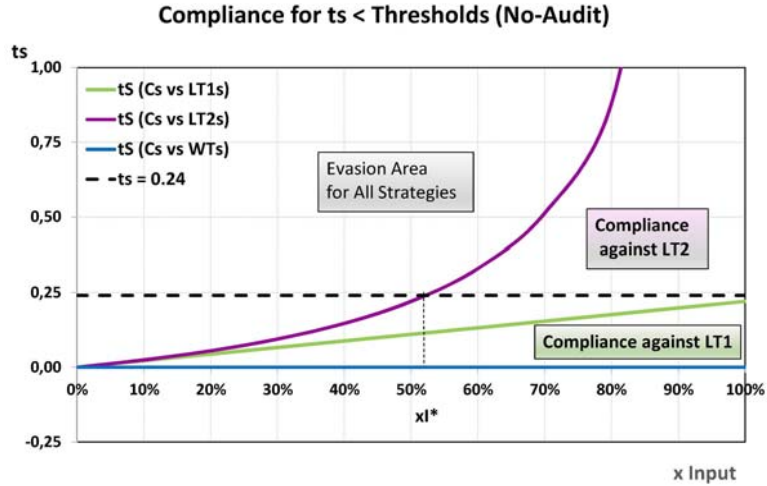

In Fig. 2. we draw the line thresholds under which the corporate tax rate $t_S$ is compatible with compliance: against $E\_LT1_S$ (green solid line), against $E\_LT2_S$ (burgundy line) or $E\_WT$ (blue x-axis). We also plot the parametric flat corporate income tax rate $t_s = 0.24$ as dashed black line. The area inducing compliance against WT evasion (on all the production chain) is simply non existent, because the tax rate should be $t_S = 0$.

As we can see, $t_s = 0.24$ can induce compliance with respect of the LT2 type of evasion, only when costs are at least 52% of the firm revenue. The same tax rate is not robust against LT1 type of evasion, i.e. when the firm is able to deduce costs anyway.

**Corollary 1** *By the intersection of each threshold line and the given corporate income tax rate, we compute the input value percentage $x_I^*$ such that for $x_I \geq x_I^*$ the Seller would comply vs evading in the three ways.*

Fixing $x_0 = 1, t_S = 0.24, v = 0.22$, and solving the inequalities by $x_I^*$ such that for $x_I \geq x_I^*$ the Seller would comply:

$$\text{for } x_I \geq x_I^* = \begin{cases} \frac{t_S x_O}{v} = \frac{0.24 \cdot 1}{0.22} = 1.0909 & \implies No-Compliance \\ \frac{t_S x_0}{v+t_S} = \frac{0.24 \cdot 1}{0.22+0.24} = 0.52174 & \implies C_S \succeq E\_LT2_S \\ \text{No-Intersection} & \implies No-Compliance \end{cases}$$

We can repeat the same reasoning and solving the three inequalities for the VAT rate, to get same threshold for $x_i$ and a symmetric plot.

**Remark 2** *In absence of audit and sanction, the only incentive to comply is the possibility to reduce net declared income with declared costs. When costs are negligible, evasion will dominate.*

### 4.1.3 Equilibria ($\gamma = 0$) _&_ T

Returning to the game

| Buyer \ Seller | $C_S$ | $E\_LT1_S$ | $E\_LT2_S$ | $E\_WT_S$ |
|---|---|---|---|---|
| $C_B$ | (1) | $\nexists$ | $\nexists$ | $\nexists$ |
| $E_B$ | $\nexists$ | (2) | (3) | (4) |

By the discussion above all four cases can be Nash Equilibria.

Under NO-Audit, the emerging Nash-Equilibrium depend essentially on the form of the production function of the chain, that we chose to omit in this treatment, but only that helps the ranking. Here we can just observe what happens in the different firm situations, discriminated by the importance of costs on firm revenue.

We can list Nash Equilibria as follows:

**Proposition 2** *(i) If the previous segments of the production chain chose evasion $\implies NE : \{E_B, E\_WT_S\}$.*
*(ii) If costs can be deduced from income tax $\implies NE : \{E_B, E\_LT1_S\}$.*
*(iii) If costs are not deductible from income tax $\implies NE : \{E_B, E\_LT2_S\}$ for $x_I < x_{I^*}$, as defined in Corollary 1.*
*(iv) If costs are not deductible from income tax $\implies NE : \{C_B, C_S\}$ for $x_I \geq x_{I^*}$, as defined in Corollary 1.*

**Proof.** By. By $(4), (5), (6)$. and Corollary $(1)$. ∎

## 4.2 Scenario A2: Taxes with Deductions under No-Audit ($\gamma = 0$) _&_ TD

We here assume:

**Assumption 1** *Some VAT invoiced consumer costs can be deducted from the Personal Income Tax. The VAT rate can as well be reduced by a percentage $\delta \in [0,1]$. The deductible share is assumed to be one-shot $\theta \in [0,1]$, where the same $\theta$ can also represent the present value of a deduction stream distributed over time.*

### 4.2.1 Payoffs $(\gamma = 0)$ _&_ $TD$

Under this scenario, payoffs change only in the compliance event (case (1)) $\{C\_D_B, C_S\}$, for the Buyer side. Payoffs in the other events are untouched.

**Event "Compliance with deductions"** $\{C\_D_B, C_S\}$

| Scenario A: NO-AUDIT $(\gamma = 0)$ |
| --- |
| Scenario A2: Taxes with Deductions $(TD)$ |
| $\{C\_D_B, C_S\} : \begin{cases} Y_B(C_B, C_S) = (1-t_B)y_B - x_O(1+\delta v) + \theta x_O(1+\delta v) \\ Y_S(C\_D_B, C_S) = (1-t_S)(y_S + x_O - x_I) \\ Y_G(C\_D_B, C_S) = t_S(y_S + x_O - x_I) + t_B y_B + v\delta x_O - \theta x_O(1+\delta v) \end{cases}$ |

### 4.2.2 Strategic Dominance $(\gamma = 0)$ _&_ $TD$

**Seller** $(TD \& \gamma = 0)$  Same as before.

**Buyer** $(TD \& \gamma = 0)$

**Proposition 3** *Buyer prefers to comply if the deduction rate is larger than a boundary*

$$C\_D_B \succeq E_B$$

*if*

$$\theta \geq \frac{v\delta}{1+v\delta} \qquad (7)$$

**Proof.** By

$$\begin{aligned} Y_B(C\_D_B) &\geq Y_B(E_B|_{s_v > 0}) \\ \theta(1+v\delta) - v\delta &\geq 0 \end{aligned} \qquad (8)$$

∎

**Remark 3** *Under No-Audit, for $\theta$ above the line represented by eq. (7), the agreement upon evasion between Buyer and Seller breaks out, and there will be a bargaining about which joint strategy to adopt.*

The algebraic sum of the buyer convenience to comply and the seller convenience to evade in every type of evasion will be computed in the Coalition Game in section.

## 5 Scenarios B: Audit and Sanctions $(\gamma = 1)$

Now all payoffs are reviewed in case of successful audit and consequent sanctions.

## 5.1 Scenario B1. Taxes with no deductions under Audit ($\gamma = 1$) _&_ T

### 5.1.1 Payoffs ($\gamma = 1$) _&_ T

| Scenario B: AUDIT ($\gamma = 1$) |
|---|
| Scenario B1: Taxes with No Deductions ($T$) |
| $\{C_B, C_S\}$ : as Scenario A1 |
| : $\begin{cases} Y_B(C_B, C_S) = (1 - t_B) y_B - x_O (1 + v) \\ Y_S(C_B, C_S) = (1 - t_S)(y_S + x_O - x_I) \\ Y_G(C_B, C_S) = t_B y_B + t_S(y_S + x_O - x_I) + v x_O \end{cases}$ |
| $\{E_B, E\_LT1_S\}$ : |
| : $\begin{cases} Y_B(E_B, E\_LT1_S) = (1 - t_B) y_B - x_O \\ Y_S(E_B, E\_LT1_S) = (1 - t_S)(y_S - x_I) - v x_I + x_O(1 - t_S(1 + s_Y)) - v(x_O - x_I)(1 + s_V) \\ Y_G(E_B, E\_LT1_S) = t_B y_B + t_S(y_S - x_I) + v x_I + x_O \cdot t_S(1 + s_Y) + v(x_O - x_I)(1 + s_V) \end{cases}$ |
| $Y_S(E_B, E\_LT2_S)$ : |
| : $\begin{cases} Y_B(E_B, E\_LT2_S) = (1 - t_B) y_B - x_O \\ Y_S(E_B, E\_LT2_S) = (1 - t_S) y_S - x_I(1 + v) + x_O(1 - t_S(1 + s_Y)) - v(x_O - x_I)(1 + s_V) \\ Y_G(E_B, E\_LT2_S) = t_B y_B + t_S y_S + v x_I + x_O \cdot t_S(1 + s_Y) + v(x_O - x_I)(1 + s_V) \end{cases}$ |
| $Y_S(E_B, E\_WT_S)$ : |
| : $\begin{cases} Y_B(C_B, C_S) = (1 - t_B) y_B - x_O \\ Y_B(C_B, C_S) = (1 - t_S) y_S - x_I + x_O(1 - t_S(1 + y_S)) - v x_O(1 + s_V) \\ Y_G(C_B, C_S) = t_B y_B + t_S y_S + t_S x_O(1 + s_Y) + x_O v(1 + s_V) \end{cases}$ |

- Plot using Italy tax rates: $y_S = 0.24, v = 0.22, y_S = s_v = 0.33$

**Fig. 3.**

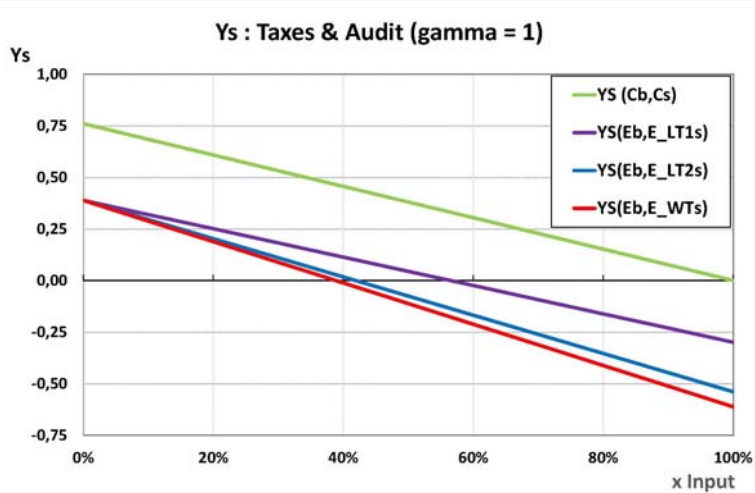

In fig.3. we plot seller net incomes $Y_S$ deriving from the four strategies $C_S$ (green line), $E\_LT1_S$ (burgundy), $E\_LT2_S$ (blue line) and $E\_WT_S$ (red line) under audit. To be noticed that this time compliance stands clearly over all others since sanctions reduce sensibly net incomes. The most penalized strategy is what was most profitable before, $E\_WT_S$, because of highest sanctions. $E\_LT2_S$ is the most rewarding evasion strategy because the ability of deducing costs from income tax is essentially not sanctionable.

Using the assumption of previous income $y_S = 0$, it's clear that evading and being caught drives net income in negative territory.

**Remark 4** *Sanctions rank progressive evasion into decreasing net income after taxes and sanctions.*

### 5.1.2 Strategic Dominance ($\gamma = 1$) _&_ $T$

As evident from the above plot, under sure audit, compliance is the dominating strategy for all agents and against all types of evasion. Moreover we assumed that the buyer cannot be caught, so he will simply follow the seller decision.

### 5.1.3 Equilibria ($\gamma = 1$) _&_ $T$

Under certainty to be caught, compliance dominates.

## 5.2 Scenario B2: Taxes with Deductions under Audit ($\gamma = 1$) _&_ $TD$

Under this scenario, compliance dominates, so auditing doesn't make any difference to the payoffs. Therefore payoffs are the same for any probability of auditing $\gamma \in [0,1]$.

# 6 Scenarios C: Uncertain Audit (Bayesian Game) $\gamma \in [0,1]$

We now enter the Bayesian game setting, where the two private agents don't know whether the third player, the Tax Authority, will audit or not. For each strategy agents compute the expected value of the two payoffs under the two extreme events Audit ($\gamma = 0$) and No-Audit weighted by the probability of auditing ($\gamma = 1$).

The novelty in this payoffs is represented by the probability $\gamma$ having a role in the thresholds.

### 6.0.1 Scenario C0. No Taxes: $NT \& (\gamma \in [0,1])$

Compliance implies no sanctions and therefore payoffs are the same as before.

## 6.1 Scenario C1: Taxes with no deductions under "Uncertain Audit" $T \& (\gamma \in [0,1])$

### 6.1.1 Payoffs $T \& \gamma \in [0,1]$

| Scenario C: Uncertain AUDIT $\gamma \in [0,1]$ |
|---|
| Scenario C1: Taxes with No Deductions $(T)$ |
| |
| $\{C_B, C_S\}|_{Bayes}$: as Scenario A1 and B1 |
| $: \begin{cases} Y_B(C_B, C_S) = (1-t_B)y_B - x_O(1+v) \\ Y_S(C_B, C_S) = (1-t_S)(y_S + x_O - x_I) \\ Y_G(C_B, C_S) = t_B y_B + t_S(y_S + x_O - x_I) + vx_O \end{cases}$ |
| $\{E_B, E\_LT1_S\}|_{Bayes}$: |
| $: \begin{cases} EY_B(E_B, E\_LT1_S) = (1-t_B)y_B - x_O \\ EY_S(E_B, E\_LT1_S) = (1-t_S)(y_S - x_I) + x_O - vx_I - \gamma x_O t_S(1+s_Y) - \gamma v(x_O - x_I)(1+s_V) \\ EY_G(E_B, E\_LT1_S) = t_B y_B + t_S(y_S - x_I) + vx_I + \gamma x_O \cdot t_S(1+s_Y) + \gamma v(x_O - x_I)(1+s_V) \end{cases}$ |
| $\{E_B, E\_LT2_S\}|_{Bayes}$: |
| $: \begin{cases} EY_B(E_B, E\_LT2_S) = (1-t_B)y_B - x_O \\ EY_S(E_B, E\_LT2_S) = (1-t_S)y_S + x_O - x_I(1+v) - \gamma x_O t_S(1+s_Y) - \gamma v(x_O - x_I)(1+s_V) \\ EY_G(E_B, E\_LT2) = t_B y_B + t_S y_S + vx_I + \gamma(x_O \cdot t_S(1+s_Y) + v(x_O - x_I)(1+s_V)) \end{cases}$ |
| $\{E_B, E\_WT_S\}| Bayes$: |
| $: \begin{cases} EY_B(E_B, E\_WT_S) = (1-t_B)y_B - x_O \\ EY_S(E_B, E\_WT_S) = (1-t_S)y_S + x_O - x_I - \gamma \cdot x_O t_S(1+y_S) - \gamma x_O v(1+s_V) \\ EY_G(E_B, E\_WT_S) = t_B y_B + t_S y_S + \gamma \cdot x_O(t_S(1+s_Y) + v(1+s_V)) \end{cases}$ |

### 6.1.2 Strategic Dominance $T$ & ($\gamma \in [0,1]$)

The procedure to define strategic dominance is the same as in Scenario A.

**Proposition 4** *In the Bayesian game, under uncertain Audit $\gamma \in [0,1]$, the Seller would prefer to comply if*

$$\text{If } t_S \leq \begin{cases} \frac{vx_I + \gamma v(x_O - x_I)(1+s_V)}{x_O - \gamma x_O(1+s_Y)} & \implies C_S \succeq E\_LT1_S \\ \frac{vx_I + \gamma(v(x_O - x_I)(1+s_V))}{(x_O - x_I) - \gamma x_O(1+s_Y)} & \implies C_S \succeq E\_LT2_S \\ \frac{\gamma \cdot x_O v(1+s_V)}{((x_O - x_I) - \gamma \cdot x_O(1+s_Y))} & \implies C_S \succeq E\_WT_S \end{cases} \quad (9)$$

**Proof.** By the incentives contraints to induce compliance against each of the three types of evasion. ∎

Threshold curves move up as the probability of audit, $\gamma$, increases from 0 to 1, creating a much larger area for compliance (the area is below the curves). For $\gamma = 0$. in Fig. 2, the strategy to evade all the production chain is the most profitable and the curve was the x-axis implying no compliance for all parameters.. As soon as $\gamma$ increases even full evasion becomes less appealing, because if caught, sanctions are very high. For $\gamma \longrightarrow 1$, curves shift to the left, and the areas below the curves include the full positive orthant leading any range of parameters compatible with compliance.

In fig.4.below, we want to compare the situation with the plot in fig. 2 (No-Audit). We consider the three compliance conditions in a same graph, fixing $\gamma = 0.15$. With respect to the No-Audit case, we note that the lines move up-left, increasing the area for compliance even for the $WT$ strategy. The thresholds $x_I^*$ determines, for which cost share, compliance becomes dominant against evasion of the considered type. The three thresholds decrease , i.e. compliance becomes profitable for firms with $x_I > x_I^*$ for each type of evasion.

Plotting parameters: $x_O = 1, v = .22, t_S = .24, s_y = s_v = 0.33$.

**Fig. 4.**

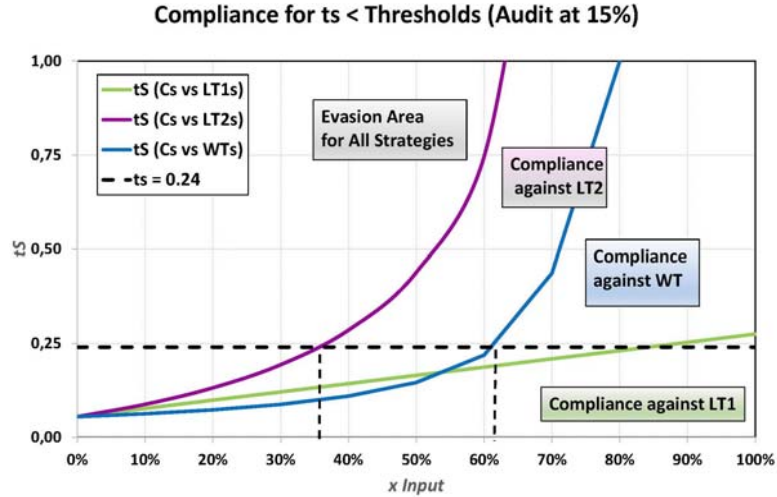

**Remark 5** *For $\gamma > 0$, the boundaries shift to the left increasing the area of compliance for lower importance of costs.*

The same reasoning applies by solving dominance for the VAT rate or for the Audit probability $\gamma$.

Plots for $v$ in this case are just symmetric to $t_S$ plots and they determine the same $x_I^*$ points as before, so we omit them.

Solving for $\gamma$ leads to additional insights.

**Proposition 5** *In the Bayesian game, under uncertain Audit $\gamma \in [0,1]$, the Seller would prefer to comply if*

$$\text{If } \gamma \geq \begin{cases} \frac{t_S x_O - v x_I}{(x_O \cdot t_S(1+s_Y) + v(x_O - x_I)(1+s_V))} & \implies C_S \succeq E\_LT1_S \\ \frac{t_S(x_O - x_I) - v x_I}{(x_O \cdot t_S(1+s_Y) + v(x_O - x_I)(1+s_V))} & \implies C_S \succeq E\_LT2_S \\ \frac{t_S(x_O - x_I)}{x_O(t_S(1+s_Y) + v(1+s_V))} & \implies C_S \succeq E\_WT_S \end{cases} \quad (10)$$

**Proof.** By the usual incentives contraints to induce compliance against each of the three types of evasion. ∎

Plotting $\gamma \geq (10)$ thresholds for $x_O = 1, v = .22, t_S = .24, s_y = s_v = 0.33$.

**Fig. 5.**
$E\_LT1$ (black), $E\_LT2$ (blue), $E\_WT$ (magenta), $t_S = 0.24$(dashed)

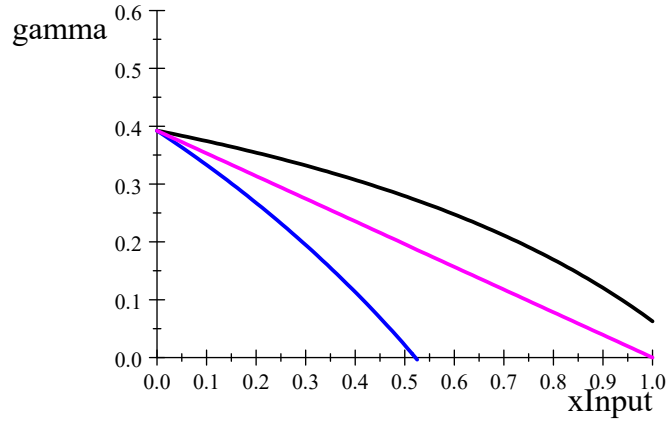

**Remark 6** *Auditing need, $\gamma$, shows the weakness of the system incentives. The three boundaries show that: (i) if the seller is able to deduce his costs (E_LT1, black line), evasion incentives are very strong so a positive $\gamma$ is needed at every level of $x_I$.. (ii) In the case of LT2 (blue line) auditing starts to be needed under the usual threshold of $x_I/x_O = 52\%$ (in this combination of parameters) when cost deductions are not important. (iii) If the seller fully evades (E_ WT, magenta line), there is need to audit him at every $x_I$ value, although with decreasing probability.*

### 6.1.3 Equilibria $(\gamma \in [0,1])$ _&_T

Equilibria are now in mixed strategy and the event $\{C_B, C_S\}|_{Bayes}$ is a Mixed Strategy Nash Equilibrium if incentive contraints apply.

## 6.2 Scenario C2: Tax Deductions under Uncertain Audit $TD$ & $(\gamma \in [0,1])$

### 6.2.1 Payoffs $TD$ & $(\gamma \in [0,1])$

Just the $\{C\_D_B, C_S\}$ event changes, payoffs in all other events are same as Scenario $C$. Moreover, since compliance gets no sanctions, auditing doesn't affect the payoffs. So this case is identical to Scenario A2.

### 6.2.2 Strategic Dominance $TD$ & $(\gamma \in [0,1])$

By comparing buyer compliance with tax discounts, to evasion strategy, compliance strongly dominates.

### 6.2.3 Equilibria $TD$ & ($\gamma \in [0,1]$)

Here, the buyer would prefer to comply, while the seller may comply only under the mentioned thresholds. The outcome is therefore unpredictable unless the threshold would be satisfied.

Therefore it is important to have the two counterparts to "cooperate" in a coalition, to decide whether the surplus gained by buyer tax allowances is able to compensate the loss of the seller and therefore the thresholds to be less binding.

## 7 Scenario D: Bayesian Coalition Game ($BCG$)

The scenario in which Buyer+Seller collude against Tax Authority is presented. The coalition agrees in scenarios with No-Taxes and Taxes, but not in the Scenarios with Tax deductions.

Along the study we have explored larger and smaller agents' advantages to diverge from collusion, depending on the cost structure of the firm or fiscal parameters, but, at last, buyer and seller have to necessarily converge on the same strategy.

To explore the preferences of the coalition, we now sum the payoffs on the principal diagonal of the previous games owned by buyer and seller. The payoffs will be listed as column elements of a game in which the row player is the private coalition (B+S) and the column player is the Tax Authority (G) auditing with probability $\gamma$.

In this way the many games are shrinked into a single one, and we can compute the expected value of each row, that corresponds to the expected value of each joint strategy with respect to the audit probability $\gamma$, provided agents "agree" on the same strategy: complying or evading.

The novelty is then to compute by which deduction rate the coalition (B+S) will agree on compliance. The question is how much surplus needs to be created by the buyer to convince the seller to comply. To do this, we consider the following game where we just consider the sum of Scenario 3 (Bayesian) payoffs for each (B+S) strategy and compute the expected value of each joint action (values in the EV column) to establish dominance between them.

| \ G | $E_\gamma V$ |
|---|---|
| (B+S) | |
| $C\_D_B + C_S$ | (1) |
| $E_B + E\_LT1_S$ | (2) |
| $E_B + E\_LT2_S$ | (3) |
| $E_B + E\_WT_S$ | (4) |

This game, labelled Bayesian Coalition Game (BCG) should be built for each scenario: No Taxes (NT), Taxes with no deductions (T), Taxes with deductions (TD).

Since in scenarios No Taxes and Taxes with no deductions there is sure agreement in the coalition, payoffs do not lead to any new result on strategic dominance. Therefore, to avoid redundancies, only the TD scenario will be considered below, where, in the compliance event, the buyer enjoys tax allowances, while in the evasion events, the seller evasion payoffs are those from the Bayesian case.

## 7.1 Scenario D2: BCG & Tax Deduction Regime ($TD\&BCG$)

### 7.1.1 Payoffs: ($TD\&BCG$)

| Scenario D: Coalition Game $\gamma \in [0,1]$ |
|---|
| Scenario D2: Taxes with Deductions ($TD$) |
| |
| $\{C\_D_B + C_S, \gamma \in [0,1]\}$ |
| $\begin{cases} EY_B(C\_D_B + C_S) + EY_S(C\_D_B, C_S) = \\ = (1-t_B)y_B - (1-\theta)x_O(1+\delta v) + (1-t_S)(y_S + x_O - x_I) \\ Y_G(C\_D_B + C_S) = t_By_B + t_S(y_S + x_O - x_I) - \theta x_O(1+\delta v) + \delta v x_O \end{cases}$ |
| $\left\{E_B\vert_{Bayes} + E\_LT1_S\vert_{Bayes}, \gamma \in [0,1]\right\}$ |
| $\begin{cases} EY_B(E_B, E\_LT1_S) + EY_S(E_B, E\_LT1_S) = \\ = (1-t_B)y_B + (1-t_S)(y_S - x_I) - vx_I - \gamma x_O t_S(1+s_Y) - \gamma v(x_O - x_I)(1+s_V) \\ EY_G(E_B, E\_LT1_S) = t_S(y_S - x_I) + t_By_B + vx_I - \gamma(x_O \cdot t_S(1+s_Y) + v(x_O - x_I)(1+s_V)) \end{cases}$ |
| $\left\{E_B\vert_{Bayes} + E\_LT2_S\vert_{Bayes}, \gamma \in [0,1]\right\}$ |
| $\begin{cases} EY_B(E_B, E\_LT2_S) + EY_S(E_B, E\_LT2_S) = \\ = (1-t_B)y_B + (1-t_S)y_S - x_I(1+v) - \gamma x_O t_S(1+s_Y) - \gamma v(x_O - x_I)(1+s_V) \\ EY_G(E_B, E\_LT2) = t_By_B + t_Sy_S + vx_I + \gamma(x_O \cdot t_S(1+s_Y) + v(x_O - x_I)(1+s_V)) \end{cases}$ |
| $\left\{E_B\vert_{Bayes} + E\_WT_S\vert_{Bayes}, \gamma \in [0,1]\right\}:$ |
| $\begin{cases} EY_B(E_B, E\_WT_S) + EY_S(E_B, E\_WT_S) = \\ = (1-t_B)y_B + (1-t_S)y_S - x_I - \gamma \cdot x_O t_S(1+s_Y) - \gamma x_O v(1+s_V) \\ EY_G(E_B, E\_WT_S) = t_By_B + t_Sy_S + \gamma \cdot x_O t_S(1+s_Y) + \gamma x_O v(1+s_V) \end{cases}$ |

### 7.1.2 Strategic Dominance ($TD\&BCG$)

As before we proceed by checking dominance between compliance with tax deductions against each of the three evasion strategies.

To check strategic dominance of compliance we set the usual three incentive constraints, i.e. inequalities between coalition compliance payoffs and each of the evasion strategies.

The inequalities can be solved by respectively by $t_S$, $v$, $\theta$ and $\gamma$. Plots are presented below for the interesting parameter constraints to achieve compliance.

**Solving Dominance Constraints for $t_S$**

**Proposition 6** *In the Coalition game, under uncertain Audit $\gamma \in [0,1]$, the coalition "Buyer + Seller" would prefer to comply*

$$\text{If } t_S \leq \begin{cases} \frac{1}{x_O(1-\gamma(1+s_Y))}\left(\theta x_O(1+v\delta) - v(\delta x_O - x_I) + \gamma \cdot v(x_O - x_I)(1+s_V)\right) \\ \qquad \Longrightarrow EY(C_B + C_S) \geq EY(E_B + E\_LT1_S) \\ \frac{1}{x_O - x_I - \gamma x_O(1+s_Y)}\left(\theta x_O(1+v\delta) - v(\delta x_O - x_I) + \gamma \cdot v(x_O - x_I)(1+s_V)\right) \\ \qquad \Longrightarrow EY(C_B + C_S) \geq EY(E_B + E\_LT2_S) \\ \frac{1}{x_O - x_I - \gamma x_O(1+s_Y)}\left(\theta x_O(1+v\delta) - v\delta x_0 + \gamma \cdot v x_O(1+s_V)\right) \\ \qquad \Longrightarrow EY(C_B + C_S) \geq EY(E_B + E\_WT_S) \end{cases}$$

**Proof.** Solving (??), (??) and (??), by $t_S$. ∎

In Fig. 6.1-3, we essentially repeat the plot in Fig. 2 (Audit with $\gamma = 0$) and in Fig. 4 (Audit with $\gamma = 15\%$) of incentive constraints for the seller, where the compliance area is below the curves. In these cases, we are considering the conditions for the coalition $(B+S)$ and draw some numerical examples for some $\gamma$ and $\theta$. Starting from $\{\gamma = 15\%, \theta = 20\%\}$ we see that the parameters are not large enough to imply full compliance. With the realistic zero audit probability ($\gamma = 0$), full compliance needs deductions for about $\theta = 35\%$.

We observe that threshold lines move up-left, increasing the area for compliance even for the $WT$ strategy. The thresholds $x_I^*$, such that for $x_I > x_I^*$ compliance becomes dominant against evasion of the considered type, decrease. The exact $\theta$ such that $x_I^* = 0$ will be determined below.

(Note for the Editor: The following three pictures can be put in a single page, or their number reduced as needed, even to zero).

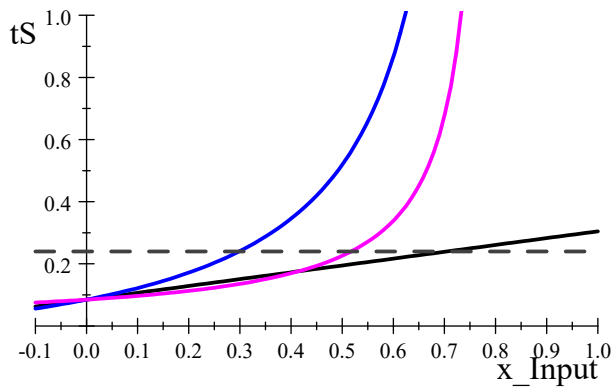

**Fig. 6.1.** $[\gamma = 0.15; \theta = 0.2]$

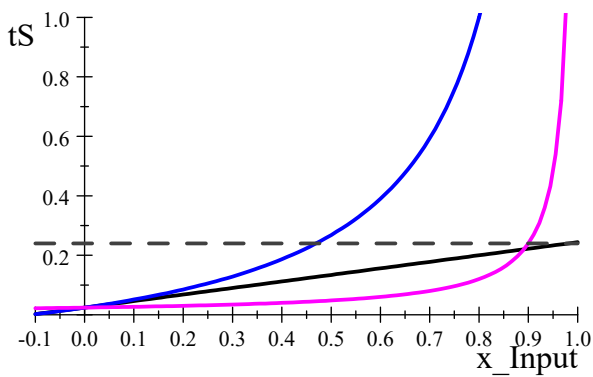

**Fig. 6.2.** $[\gamma = 0\ ; \theta = 0.2]$

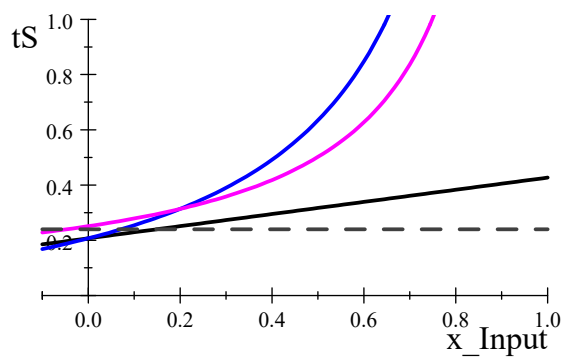

**Fig. 6.3.** $[\gamma = 0; \theta = 0.35]$

The same reasoning applies by solving dominance inequalities for the VAT rate, for the deduction rate $\theta$ or for the Audit probability $\gamma$.

**Solving Dominance Constraints for $\gamma$**

**Proposition 7**

$$\text{If } \gamma \geq \begin{cases} \frac{t_S x_O + v(\delta x_O - x_I) - \theta x_O (1+v\delta)}{v(x_O - x_I)(1+s_V) + t_S x_O (1+s_Y)} & \implies E(C_B + C_S) \geq E(E_B + E\_LT1_S) \\ \frac{t_S(x_O - x_I) + v(\delta x_O - x_I) - \theta x_O (1+v\delta)}{v(x_O - x_I)(1+s_V) + t_S x_O (1+s_Y)} & \implies E(C_B + C_S) \geq E(E_B + E\_LT2_S) \\ \frac{t_S(x_O - x_I) + v\delta x_O - \theta x_O (1+v\delta)}{v x_O (1+s_V) + t_S x_O (1+s_Y)} & \implies E(C_B + C_S) \geq E(E_B + E\_WT_S) \end{cases} \quad (11)$$

**Proof.** By solving (??), (??) and (??), by $\gamma$. ∎

Let's compare this situation with the situation in Fig. 5., where the seller was deciding alone.

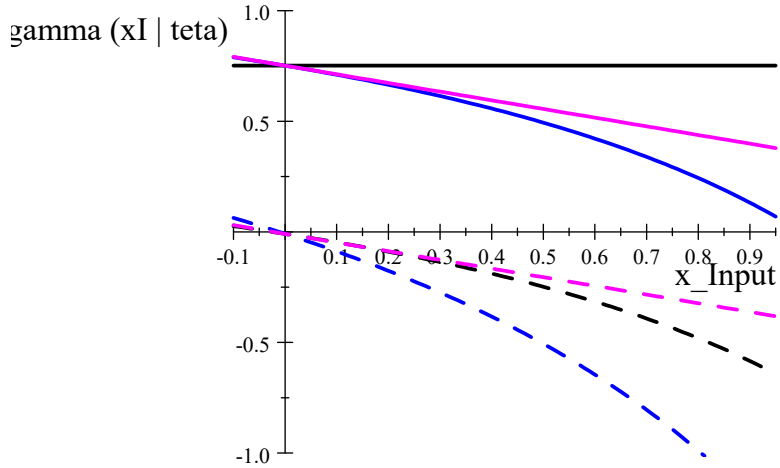

**Fig.7.1..** For each constraint $\theta = 0$ (solid); 0.35 (dash).

LT1 (Black) LT2 (Blue), WT (Magenta)

In fig. 7, we show how the threshold curves against each evasion strategy (11) shift down as $\theta$ increases from 0 to 1. To do that, we draw the three threshold curves (LT1, LT2, WT) in different colors for $\theta = 0$ (solid) and 0.35 (dash).

Not very surprisingly, the most profitable evasion strategy is LT1, where the firm manages to deduce $x_I$ from previous revenue, reducing net imposable income. When tax allowances are not admitted, i.e. $\theta = 0$, audit frequency $\gamma$ required against LT1 evasion is the maximum. As allowances increase $\theta > 0$, $\gamma$ required for compliance decreases and relatively more steeply for the high costs firms, because, as previously stated, they're more keen to comply because of large sunk VAT on a larger $x_I$. We notice that for $\theta > 0.35$, compliance is achieved for every $x_I \in [0, 1]$.

We now look for the exact tax allowances rate $\theta$, able to realize compliance in absence of auditing. In equations (11), let's compute $\theta_i^*(\gamma = 0)$ against each strategy and along any $x_I \in [0, 1]$.

**Corollary 2** *The* tax allowance fully compensating audit probability *for each strategy $i = LT1, LT2, WT$, to ensure compliance for any $x_I \in [0, 1]$.is $\theta_i^*(\gamma = 0)$*

$$\begin{cases} \theta_{LT1}^*(x_I \mid \gamma = 0) = \frac{t_S x_O + v(\delta x_O - x_I)}{x_O(1+v\delta)} & \implies E(C_B + C_S) \geq E(E_B + E\_LT1_S) \\ \theta_{LT2}^*(x_I \mid \gamma = 0) = \frac{t_S(x_O - x_I) + v(\delta x_O - x_I)}{x_O(1+v\delta)} & \implies E(C_B + C_S) \geq E(E_B + E\_LT2_S) \\ \theta_{WT}^*(x_I \mid \gamma = 0) = \frac{t_S(x_O - x_I) + v\delta x_O}{x_O(1+v\delta)} & \implies E(C_B + C_S) \geq E(E_B + E\_WT_S) \end{cases} \quad (12)$$

Let's plot the three constraints (12) as functions of the cost structure $x_I$.

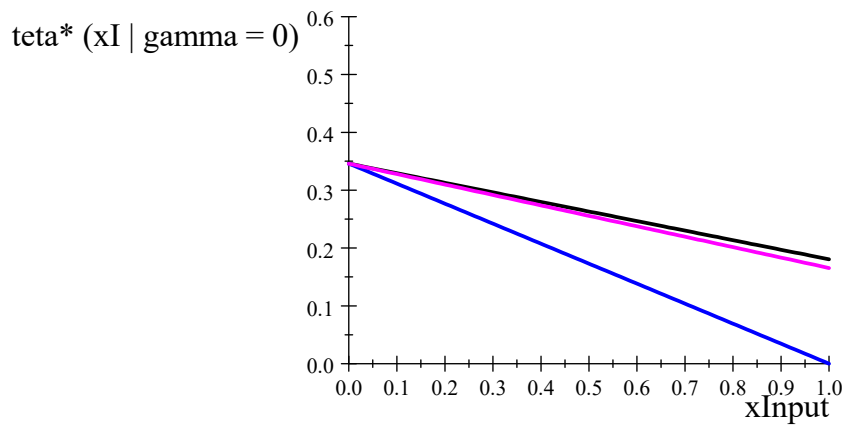

**Fig. 7.2.** $E\_LT1$ (black), $E\_LT2$ (blue), $E\_WT$ (magenta), $t_S = 0.24$(dashed)

In fig. 7.2., we observe that the incentice contraints are linear functions in $x_I$ with same intercept and different slopes for each strategy. So $\theta_i^*(x_I = 0|\gamma = 0) \approx 0.35$ should ensure compliance with no need of auditing for all costs structure. The slope is constant for all seller evasion strategies because the slope measures dependence on sanctions that is the same in all evasion strategies in this example. On the contrary, the intercept measuring the convenience to evade, is different for the different strategies.

Fig. 7.2. shows that as input costs increase, the rate of allowances decreases less steeply against evasions that deduce costs $(LT1, WT)$, symmetrically to what was happening in the functions $(11)$. This is because where costs are not reducing net income ($E\_LT1_S$ and $E\_WT_S$ strategies), the surplus that the buyer has to bring to the coalition to compensate for the seller loss in compliance (expressed by $\theta$), has to be larger.

In fig. 7.3. below, we plot the trade-off between the rate of allowances $\theta$ and the probability of audit $\gamma$. In this risk neutral setting, the trade-off is linear. Intercepts are higher, because more intervention is needed against more convenient evasion strategies.

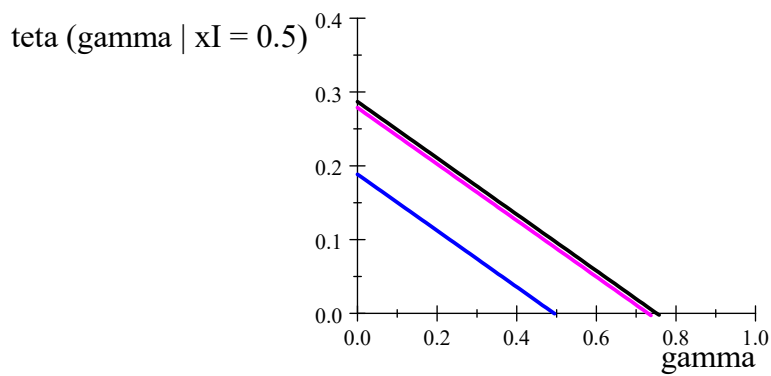

**Fig. 7.3.**

Same results as above can be reached by solving the three inequalities by $\theta$.

### 7.1.3 Coalition Bayesian Equilibrium

In this stylized setting,

**Proposition 8** *A Bayesian Equilibrium with Tax Compliance exists if the Tax-Authority determines fiscal parameters respecting the binding constraints* (??), (??) *and* (??).

The number of constraints is smaller than the fiscal parameters, so the Tax authority should decide which variables to be treated as endogenous. For example a $\theta$ valid against all evasions, as a function of a desired $\gamma$, given $t_S, t_B, v, s_Y, s_V$.

### 7.1.4 Tax Revenue Losses due to Tax Allowances

How much is the revenue loss on the Government revenue due to tax allowances in this setting?

In this stylized model the answer is very straightforward.

Let's compare full compliance government revenue and the government revenue due to tax deductions:

$$\begin{aligned} Y_G\left(C_B, C_S\right) &= t_B y_B + t_S \left(y_S + x_O - x_I\right) + v x_O \\ Y_G\left(C\_D_B + C_S\right) &= t_B y_B + t_S \left(y_S + x_O - x_I\right) + v x_O - \theta x_O \left(1 + v\right) \end{aligned}$$

The latter has a loss of $-\theta x_O \left(1 + v\right)$, that, in this context $(\theta = 0.38, v = 0.22)$, is about $-46\%$ of missed tax revenue.

To apply it empirically, we need to evaluate it just on the specific sectors where tax allowances are applied, that are normally the ones in which evasion is very high and the consequent frequency of audit very expensive.

## 8 Conclusions

We dealt with the design of tax compliance incentives able to break the convenience of the evasion by the private agents coalition (B+S) against the government G, transforming it in a virtuous coalition between the buyer and the Tax Authority (B+G) able to force sellers to comply. The success in breaking the VAT evasion agreement, although costly, has the double result to recover VAT and income tax revenue in sectors where evasion would be proportionally more expensive.

The strategic approach specifies, for each tax policy scenario and against some different types of evasion, the constraints on endogenous tax parameters necessary to obtain the strategic dominance of compliance. It also measures the trade-off between tax allowances and Tax Authority audit with respect to a stylized cost structure of the firm (input/ouput value ratio), emerging as a crucial discriminating variable in determining evasion, being also a proxy for the length of the production chain.

The strongest driver for evasion is the collusion between the buyer and the final seller. The longer the chain, the more VAT is accredited to the final seller that is therefore less willing to lose the sunk VAT payments. The larger the importance of sunk costs, the smaller the gain from evading on corporate income tax and the last VAT tranche. In particular, the model shows that there exists a threshold on the input to output values ratio $(x_I/x_O)$, depending on the given tax rates, over which compliance is spontaneous even in absence of audit. This is why, for sectors having ratios under that threshold, it is optimal to establish tax allowances for the buyer, to have him bargaining seller compliance.

Summarizing, the model computes strategic dominance constraints on fiscal parameters, to reach spontaneous compliance by the single agents in absence and in presence of a random audit. Same results are computed for the joint coalition when the buyer, in order to enjoy the tax allowances, may bargain some surplus transfer to the seller. In this case, compliance is achieved with more freedom on tax parameters and a much lower need of auditing.

The contribution of this model is to determine constrained ranges, for tax rates treated as endogenous variables, that can be easily applied to real data for precise tax design simulations.